# A Symbol-Based Estimation Technique for Inter-vehicular Communication Performance Optimization.

Aymen SASSI,
Faiza CHARFI, Lotfi KAMOUN
National Engineering School of Sfax,
University of Sfax, Lab of Electronics and Information
Technology Sfax, Tunisia
asassi.th@gmail.com, faiza.charfi@gmail.com,
Lotfi.Kamoun@isecs.rnu.tn

Yassin ELHILLALI, Atika RIVENQ
University of Valenciennes and Hainaut Cambresis
Institute of Electronics Microelectronics and Nanotechnology
Valenciennes, France
yassin.elhillali@univ-valenciennes.fr,
menhaj@univ-valenciennes.fr

*Abstract*— The aim of this paper is to enhance the quality of Orthogonal Frequency Division Multiplexing OFDM estimation in dedicated vehicular communication transmission V2X networks. Wireless Access in Vehicular Environment WAVE as also known IEEE 802.11p represents the standard for these networks. Developing a reliable inter-vehicular V2X communication has to focus on optimizing its real performances. In this work, we studied the fact that WAVE transmission uses the channel characteristics designed for indoor and stationary communication terminals in IEEE 802.11a. In this paper, we propose an approach to overcome this mobility problem of terminal communication. The considered solution consists in using pilot estimation technique to reduce the high bit error rate. First, we highlight the impact of rearranging the pilot symbol positions on the quality of transmission QoT. Second, we try to overcome one of the PHY layer estimation constraints by adding two new pilot symbols. By considering pilot symbol aided channel estimation at the transmitter, we focus on Least Square LS and Minimum Mean Square Error MMSE channel estimation on the receiver. A range of simulations is carried out according to ratio between the Bit Error Rate BER and the Signal to Noise Ratio SNR. We demonstrate that rearranging pilot pattern can offer better results than standardized ones. Furthermore, we prove that adding pilots symbols can provide the best performances.

## 1. INTRODUCTION

Intelligent Transport Systems ITS are one of the major research topics which are in a large expansion. This expansion is the result of the huge development of both the number of transport systems and the users' need for services namely collision avoidance, clear congestion and e-traffic. The communication between vehicle to vehicle V2V and or vehicle to infrastructure V2I are one of those important needs for safety applications in ITS. The IEEE community has conducted a variety of studies to standardize the V2X communication through IEEE 802.11p. It inherits medium access control MAC layer specifications from the IEEE 802.11e and physical PHY layer specifications from the IEEE 802.11a using its Orthogonal Frequency Division Multiplexing OFDM modulation. This modulation is used for the high data rate that it can offer as well as its adequate performance in selective frequency channels. Due to mobility, the fast-changing specification stat of radio channel specification can be considered as one of the major disadvantages in the 802.11p transmission. Therefore we need to adopt an efficient OFDM channel estimation based on coherent detection and decoding. To achieve this purpose, two paramount orientations have been proposed. The first one considers training reference signal in the transmitter and the receiver while the second considers the design of the best tracking ability and also a low complexity rate.

The objective of this study is to enhance the performance of the V2X communication; using a combined estimation method to reach our goal. First, we used a based pilot estimation technique to make the channel estimation more robust. The challenge that we focused on is to rearrange the pilot subcarrier positions. Consequently, a little enhancement of the bit error rate is obtained. Second, we added then two more pilot subcarriers and we proved that this method had yielded better results than the first one. We combined this first estimation method with the results based on comparison between MMSE and LSE estimations.

The rest of the paper is structured as follows: Section 2 is about an overview of the WAVE related literature. Section 3 deals with the implementation of the transmission system process. Section 4 treats based symbol estimation in OFDM system: it introduces the main techniques used in our approach. In section 5 titled a based pilot symbol 802.11p optimization; we will discuss the proposed enhancement pattern and the new added pilot symbol estimation simulation and results that are carried out in this work. Section 6 includes the conclusions and perspective works.

## 2. RELATED LITERATURE

One of the most important constraints in V2X communication networks is the mobility of the terminals which is the main cause behind the transmission weakening. This transmission fading can be caused by a range of factors among which we can note the reflected transmitted signals on spatial obstacles between transmitter and receiver. Numerous studies are being conducted to try to overcome this fading.

The work in [1] highlights the OFDM transmission performance evaluation in V2X Communication. The authors have showed the influence of the mobility on the BER. They have also proved that whenever we upgrade the number of bits to be transmitted per frame, the QoT decreases. In few OFDM systems literature studies, the main method used for error variance estimation is based on statistical approaches such as the Cramer-Rao Bound CRB method. Indeed the authors in [2] and [3] have used the CRB for frequency estimation error variance. Barry and al. in [2] have considered the problem of time synchronization. They have focused on the training sequences in digital magnetic recording systems. In their work, a more general acquisition technique was introduced: training symbols are arbitrarily placed throughout the transmission. A split-preamble arrangement is proposed to minimize the CRB and the square estimation error. The split-preamble strategy consists in dividing the training symbols in two halves: the first one is placed at the beginning of the sector while the second half is placed at its end. One of the most popular algorithms of the pilot-aided estimation schemes is the method proposed by Schmidl in [3]. This algorithm provides a simple estimator for symbol timing offset and frequency offset. The work in [4] has tried to introduce several frequency offset estimation schemes. The authors have developed a pilot-aided synchronization to avoid frequency offset in vehicular environment. They have used the correlation between preamble sequence types. The main results of their technique were to reduce the Mean Square Error MSE. The authors in [5] have considered PSK on Rayleigh fading channel to conclude that the loss due to the imperfect channel estimation depends exclusively on the signal of noise ratio. The presented approach, which can resist to both narrow-band and spread spectrum communication, can also introduce an analytical optimization of pilot symbol spacing. Wu and al. have introduced in [6] a novel semi-blind inter carrier interference ICI equalization scheme for OFDM. Their approach has used a Joint Multiple Matrix Diagonalization JMMD algorithm. Through Monte Carlo simulations using OFDM modulations (QPSK and 16 QAM), the authors have proved that their proposed method can achieve a significantly better performance with Symbol Error Rate SER reduction in different orders-of-magnitude. The work in [7] has studied the analytical bit error performance for pilot symbol aided coherence. They have conducted their studies forward of WCDMA systems with Rayleigh fading channel model. As a result of their simulations, the authors showed the existence of an optimal value of M, where M is the time duration in which one pilot symbol is inserted. This M value can enhance the relative performance loss in ideal coherent demodulation. He and al. have used in [8] a new technique based on pilot pattern optimized criteria. They have shown that the proposed approach can give better values for both BER and MSE. Monitored estimation via the pilot aided technique has been used in [9] to perform a coherent detection in a nonselective frequency fading channel. Nevertheless, the authors in [10] have proposed to use the interpolation as it can be performed using Discrete Fourier transform DFT-based techniques which provide an efficient tradeoff between complexity and performance. The authors in [11] have used the channel frequency correlation for low-rank channel estimation for OFDM systems. This method has been proposed despite its poor performance with non-spaced sample channels. They have shown that their technique is robust against noise and possible channel changes. Coleri and al. in [13] have based their research on channel estimation techniques based on pilot arrangement in OFDM systems. They summarized a global review of different pilot based channel estimators and they proved that Comb-type pilot estimation pattern with low- pass interpolation gives better results than other similar algorithms. The work in [14] concerns a Pilot-Based Channel Estimation for OFDM systems. The authors in this study have used a monitored subspace delay. They propose to insert an intermediate step between the two essential ones in OFDM systems estimation in fast changing fading channel. Their proposed technique will increase the exactness in pilot subcarriers over the considerable benefits that it produces. Leus and al. in [15] have focused their paper on Multiple-input multiple-output MIMO for OFDM systems with specific channel parameters. The equalization technique proposed, Per-Tone Equalization PTEQ, is based on changing the filtering operations of MIMO channel and the sliding FFT. They have introduced this technique as an interesting alternative to the Time-domain Equalization TEQ one. Zemen and al. have centered their research in the vehicular communication domain. They have first proposed iterative channel estimation based on 2D subspace spanned by generalized discrete prolate spheroidal sequences. They have also designed a backward compatible modification in WAVE pilot pattern. The task of this technique is to reduce the number of iterations sufficient for convergence and to enhance the implementation complexity. Emad and al. in [16] have tried in their work to study the performance of the Continuous Phase Modulation based on the OFDM systems CPM-OFDM. The authors have considered and compared in their researches two types of equalization the Zero Forcing and the MMSE equalization for the performance evaluation. They have shown that the CPM based Single Carrier Frequency Domain Equalization CPM-SC-FDE proposed can enhance the CPM-OFDM systems by about 5dB. A new estimation method was studied in [17]. The author have proposed to use a high power impulse sample and several zero samples at the end of a zero padded OFDM symbol block, in order to estimate the Channel Impulse Response CIR. The authors have focused their main topic in Impulse Postfix OFDM system IP-OFDM. They have proposed an adaptive IP scheme to enhance the BER performance of IP-OFDM systems by adjusting adaptively the IP length depending on the CIR. They have shown over various simulation results that their technique can achieve up to 2 dB performance enhancement.

Outdoor and mobility transmissions in OFDM can represent one of the major sources of fading because their fast changing channels characteristics. Hence, the overcoming of this degradation has become the biggest issue in several studies across literature. Taking into account the studies mentioned above and in view of enhancing the V2X communication, we propose an estimation methods based on pilots symbols and combined with square estimation methods. The section that follows will introduce the proposed techniques and will recapitulate first the impact of both training symbol estimation and LSE / MMSE estimation.

## 3. IMPLEMENTATION OF THE SYSTEM MODEL

The OFDM transmission system based on pilot's estimation channel is given in Fig.1. In this section, we will introduce briefly the Matlab implementation of this used system.
First, binary data are generated with a random source component. Generated data are then coded. Coding data consists in transferring the bits to be transmitted to the scrambler in order to avoid long successive bit streams of 0 and 1. Scrambled data are then coded in the convolutional encoder according to a coding rate R=1/2. This operation will also help to avoid the Inter-Symbol Interference ISI. The coded bits stream are used in the puncturing element to obtain the upper coding rates R=2/3 or R=3/4. This operation consists in omitting some coded bits according to those two bit sequences "1110" and "110101" alternately for R=2/3 and R=3/4. The output bits from the puncturing are interleaved using two permutation methods within one OFDM symbol. The first permutation ensures that two successive bits are coded in two non-adjacent subcarriers. The second permutation allows that two successive bits are represented alternately in the most and least significant bits of the used constellation. The interleaved bits are then modulated using BPSK, QPSK, 16 QAM and 64 QAM as mandatory modulation formats. The inverse fast Fourier transfer IFFT is used to convert the serial set stream in frequency domain to a parallel stream in time domain. This step places the complex symbols associated with different constellation points on subcarriers. The IEEE 802.11p uses 52 subcarriers which are exploited as follows: 48 are data subcarriers and 4 are pilot subcarriers. To avoid the ISI, a cyclic prefix CP is inserted before each OFDM symbol to overcome the possible multipath problems. It consists in copying the end of OFDM symbol at the beginning of the following symbol.

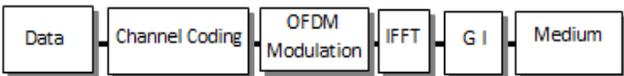

Fig. 1. OFDM transmission system.

## 4. SYMBOL-BASED ESTIMATION IN OFDM SYSTEM

In time-varying channels, estimation can point up major transmission details. It can also represent an efficient technique to solve high error rates. In this section, we underline the different algorithms dedicated to channel estimation in V2X communication. IEEE 802.11p standard uses two different types of symbol assisted estimations. Bloc-type pilot estimation is the first one used under the presumption of indoor fading channels. It consists in integrating periodically pilot tones in all subcarriers of the OFDM symbols. The second type is called Comb-type pilot estimation and unlike the bloc-type pilots, it consists in adding pilot tones in a specific sub-carrier position in each OFDM symbols.

### 4.1 Bloc-Type pilot estimation

The bloc-type pilot estimation method also called Training sequence is illustrated through a periodic sequence of symbols transmitted at the beginning of each OFDM symbol as preamble field. Their aim consists in estimating the channel state. These types of symbols are inserted on the N useful subcarriers NSc Fig 2. Green columns show the allocation of all subcarriers at one transmitting time. They are composed of ten Short Training Symbols STS given by (1) and two Long Training Symbols LTS given by (2).

$$r_{Short}(t) = w_{TShort}(t) \sum_{k=-N_{ST}/2}^{N_{ST}/2} S_k \exp(j2\pi k\Delta_F t) \quad (1)$$

$$r_{Long}(t) = w_{TLong}(t) \sum_{k=-N_{ST}/2}^{N_{ST}/2} L_k \exp(j2\pi k\Delta_F (t - T_{GI2})) \quad (2)$$

Where $N_{ST}$ denote the total number of used subcarrier and $\Delta_F$ denote the subcarrier frequency spacing. $L_k$ and $S_k$ are respectively the long training sequence and the short training sequence. Denote by $w_{TShort}$ (t) and $w_{TLong}$(t), the time-windowing function which depend on the value of the duration parameter T.

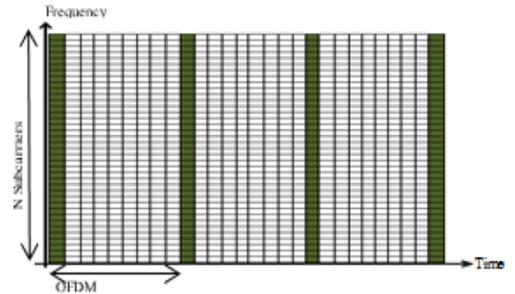

Fig. 2. Training Sequence placement.

With this kind of estimators, the receivers can decode the received data while waiting for the arrival of the next Preamble field using two estimation algorithms: Least Square LS estimation or the Minimum Mean-Square Error MMSE estimation. The least square algorithm which consists in reducing the mismatching rate between the received data and their expected values without any information about the channel statistics is used. This algorithm offers a development with a very low complexity rate but it gives rise to an important disadvantage: a high mean-square error. LS estimator is defined so as to minimize the parameter $(\overline{Y} - \underline{X}\overline{H})^H (\overline{Y} - \underline{X}\overline{H})$, where $(.)^H$ denotes the conjugate transposed operation. It is shown that the LS estimator of $\overline{H}$ is given by [18]:

$$\hat{H}_{LS} = \underline{X}^{-1}\overline{H}$$

$$= [\left(\frac{X_k}{Y_k}\right)]^T \ (k = 0,1,2,\ldots\ldots N-1)$$

Where $\overline{X}$ gives the pilot signal and $\overline{Y}$ the received signal.

In order to overcome the LS disadvantage, the MMSE method is used to minimize the mean-square error under a low SNR values. The main idea of MMSE estimation is to consider the channel frequency response and the added white noise on it while limiting the equalization coefficient in fading channels. Although the MMSE offers better results than the LS estimation method, it has an important drawback: it brings about a high algorithmic complexity. It is calculated as follows:

$$\hat{H}_{MMSE} =$$

$$= \underline{F}[(\underline{F}^H \underline{X}^H)^{-1} \underline{R}_{gg}^{-1} \sigma_N^2 + XF]^{-1}\overline{Y}$$
$$= \underline{F}\underline{R}_{gg}[(\underline{F}^H \underline{X}^H \underline{X}\underline{F})^{-1} \sigma_N^2 + \underline{R}_{gg}]\underline{F}^{-1}\hat{H}_{ls}$$

Where $\underline{R}_{gg}$ denotes the auto-covariance matrix of $\overline{g}$ and $\sigma_N^2$ denotes the variance $E\{|\overline{N}|^2\}$.

In this section, we try to underline the effect of the two previous estimation algorithms on our IEEE 802.11p OFDM transmission system. Fig 3a and 3b compare the transmission performance schemes for two modulation types in AWGN channel model. The simulation scenario defines speed to 50 Km/h and 4,000,000 bits to be transmitted.

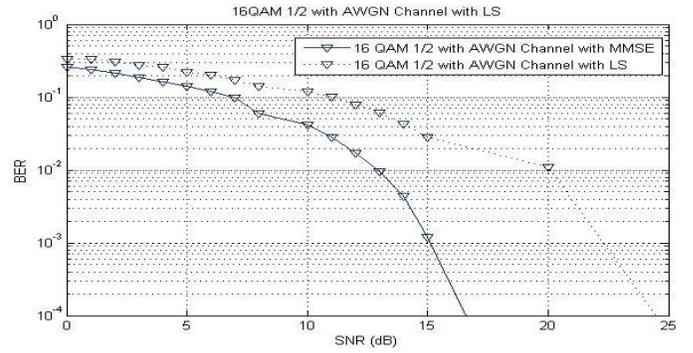

Fig. 3a. BER versus SNR for MMSE and LS channel estimators in 16 QAM modulation

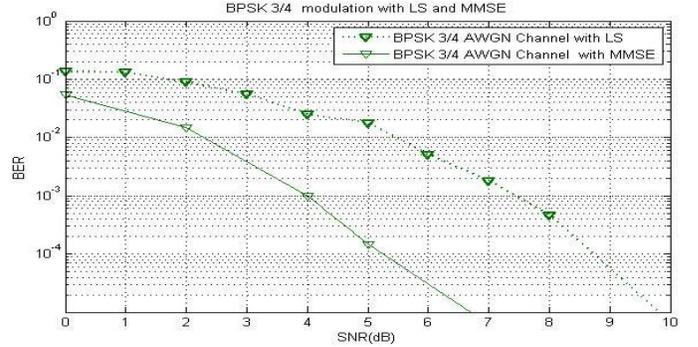

Fig. 3b. BER versus SNR for MMSE and LS channel estimators BPSK modulation

Fig. 3a and 3b show the BER versus SNR with the 16QAM modulation and BPSK modulation in AWGN channel model condition. These two plots illustrate a comparison between MMSE and LS estimation algorithms. The dashed curves describe the LS estimation technique while the continuous curves describe the MMSE estimation algorithms. The results shown in these two plots denote that the continuous curves of MMSE estimation algorithms outperform the LS estimation ones especially when SNR takes higher values. The MMSE estimator still has a major drawback; its high complexity values.

4.2 Comb-Type Pilot Channel Estimation:

To avoid fast fading transmission channels, the second type of pilot assisted estimations is transmitted during all transmission time. Indeed, Comb-type pilot symbols given by (3) are inserted uniformly in four dedicated subcarriers fixed by the 802.11p standard according to a fixed pattern Fig 4. We subsequently use an interpolation to estimate the medium for data transmission. In Fig 3, red lines show the allocation of the four pilot subcarriers in positions fixed by the IEEE 802.11p standard at -21, -7, 7, 21 positions.

P–26, 26 = {0, 0, 0, 0, 0, 1, 0, 0, 0, 0, 0, 0, 0, 0, 0, 0, 0, 0, 0, 1, 0, 0, 0, 0, 0, 0, 0, 0, 0, 0, 0, 0, 0, 1, 0, 0, 0, 0, 0, 0, 0, 0, 0, 0, 0, 0, 0, 0, –1, 0, 0, 0, 0, 0}

(3)

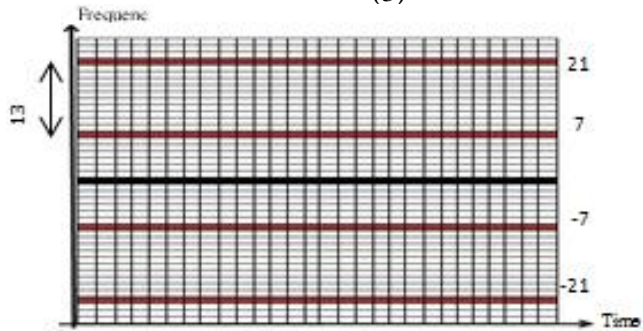

Fig. 4 Pilot Symbol insertion

## 5. A pilot-based symbol 802.11p Optimization

After compiling some references in the state of art and based on results found in [1], we will try to propose a solution for the IEEE 802.11p performance enhancement. Our proposed approach is based on a channel estimation with pilot symbols. Two sets of scenarios are performed: In the first one, we highlight the impact of rearranged position of the pilot symbol on the transmission quality QoT [12]. Second, we try to illustrate the impact of adding two new pilot symbols in the transmission process.

### 5.1 Optimized pilot position method

In this part, we have performed a bit error rate variation according to signal to noise ratio focusing on the effect of changing the pilot symbol positions in 8 modulation types. The scenario specifications consider the transmission of 4,000,000 bits on an AWGN channel model with a vehicle speed fixed at V= 50 Km/h. For the initial case, the channel estimation was set according to an irregular spacing between subcarriers schema. The spacing between the pilot subcarriers is as follows: [-21, -7, 7, 21]. It influences the number of non-estimated subcarriers (maximum 13 in Fig. 3).

To improve the transmission performance, we propose to reduce the number of these non-estimated subcarriers. The pilot sub-carriers are repositioned with a regular spacing pattern of 8 Fig.5. This proposed technique has the advantage of reducing the number of non-estimated subcarriers to 8.

In the light of studies in the previous section, it turns out that the MMSE is better than LS estimation. For that reason we have used this estimator in our scenarios.

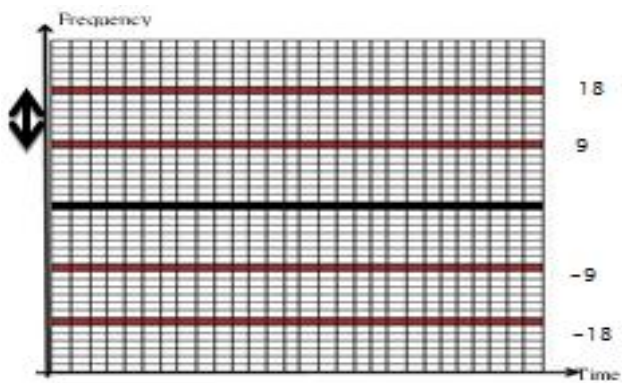

Fig.5- Rearranged pilot subcarrier positions

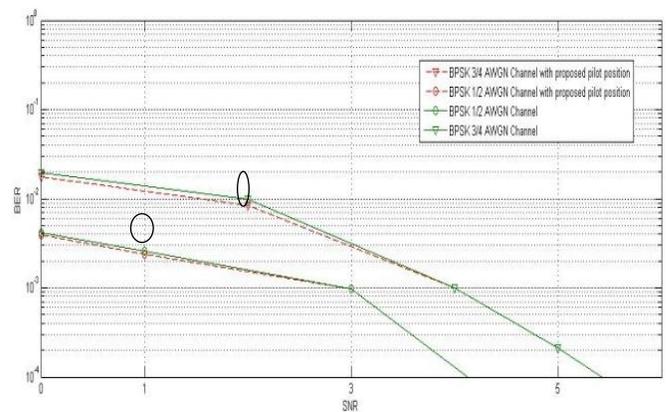

Fig. 5a, Standardized estimation Vs New Pilot position estimation for BPSK

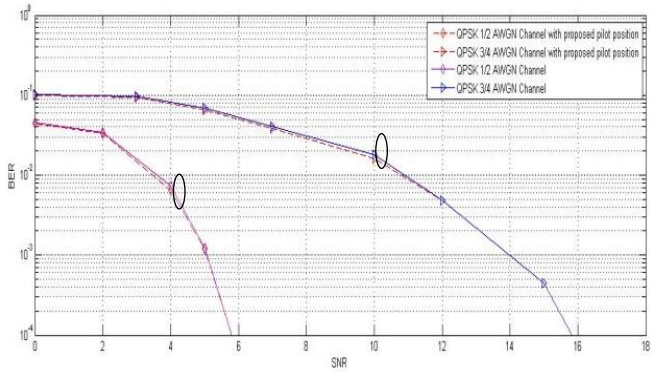

Fig. 5b, Standardized estimation Vs New Pilot position estimation for QPSK

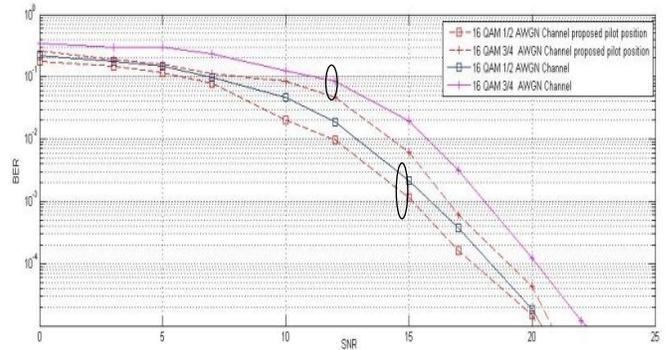

Fig. 5c Standardized estimation Vs New Pilot position estimation for 16 QA M

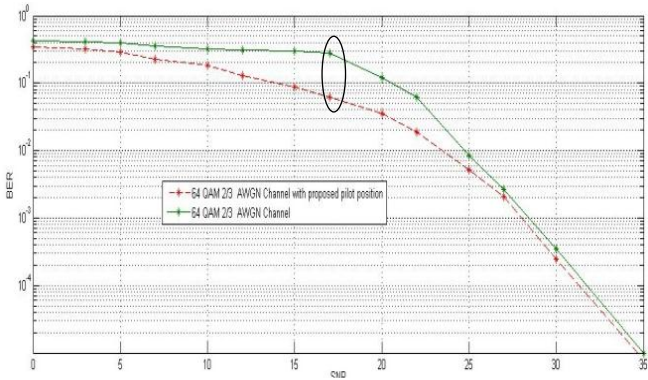

Fig. 5d Standardized estimation Vs New Pilot position estimation for 64 QAM 2/3

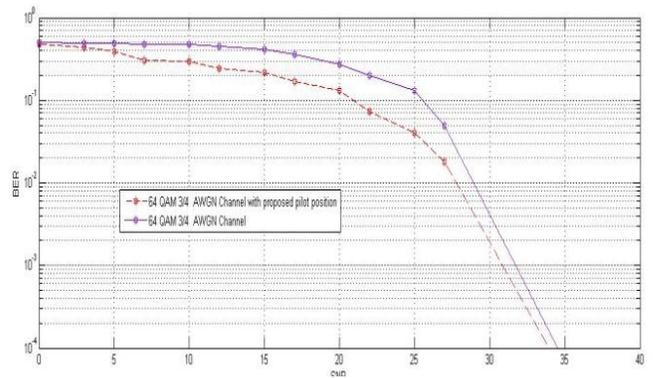

Fig. 5 e Standardized estimation Vs New Pilot position estimation for 64 QAM ¾

With our proposed rearranged pilot position technique and according to the Fig. 5(a, b, c, d, e), we can denote that moving the pilot subcarriers can provide better simulation results by reducing the series of the non-estimated subcarriers. We can also prove that the channel state will be more predictable and by the way enhance the bits transmission.

### 5.2 Optimized pilots number method

In this part, we try to study two scenario sets. For this purpose, we have first made a confrontation between a case where we consider the proposed and standardized pilot positions defined above. Then, we have proposed to append two supplemented pilot subcarriers.

Based on the fact that the number of used subcarrier is 52 and the fact that IFFT elements used can offer 64 subcarriers, we have proposed to devote 2 new subcarriers to pilot symbols as we have 8 non used ones.

For these simulations, we have used the same bits number to be transmitted 4,000,000. Transmission will also be with a speed V= 50 Km/h considering the AWGN channel model for the 8 modulation types with the MMSE estimation method.

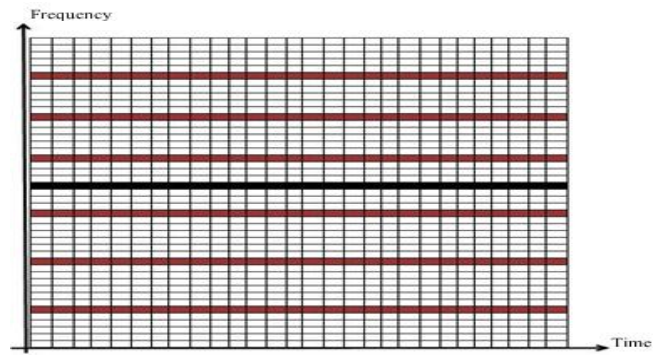

Fig. 6 Transmission media with two added subcarrier

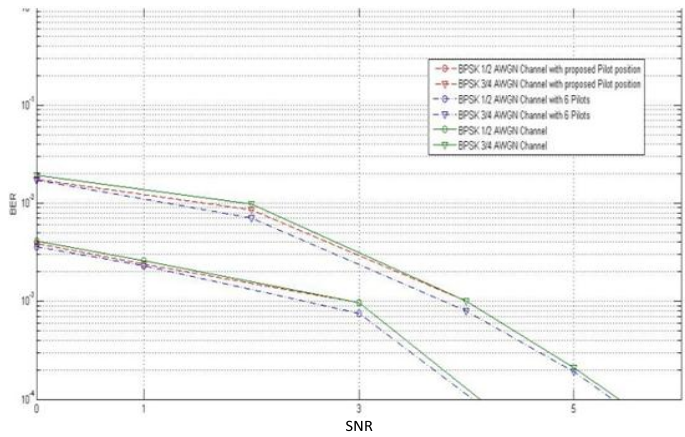

Fig. 6. a. Confrontation between added pilot subcarrier and repositioned pilot subcarrier for BPSK modulation

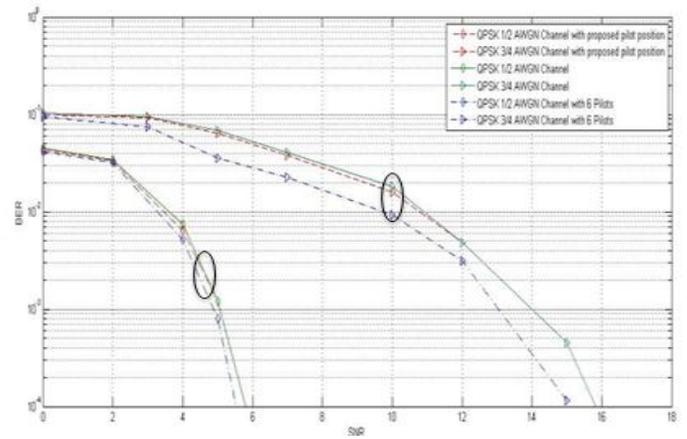

Fig. 6. b Confrontation between added pilot subcarrier and repositioned pilot subcarrier for QPSK modulation.

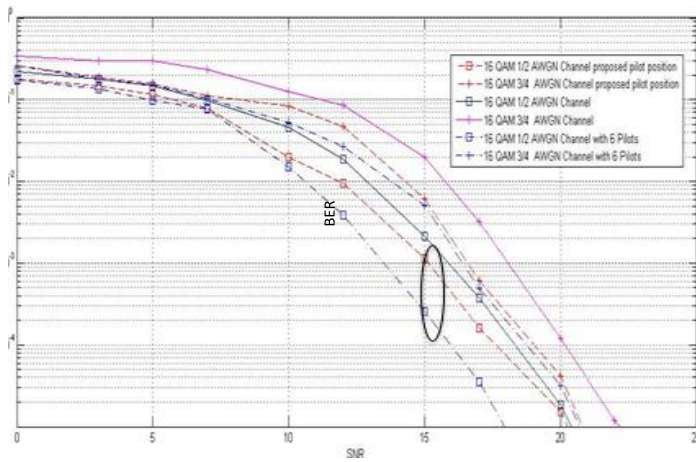

Fig. 6. c Confrontation between added pilot subcarrier and repositioned pilot subcarrier for 46QAM modulation.

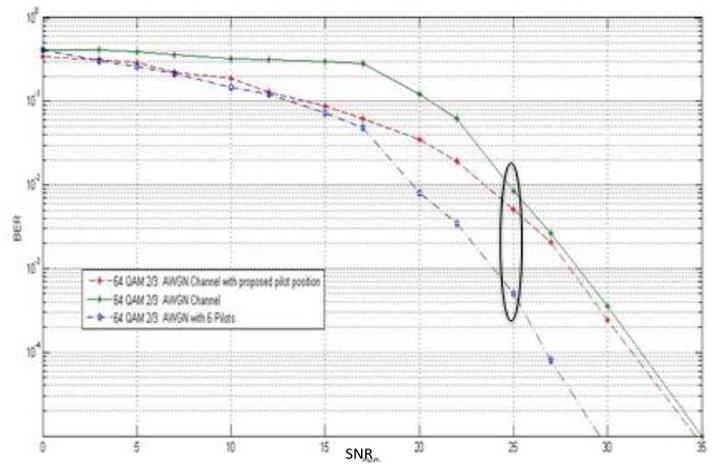

Fig. 6. d Confrontation between added pilot subcarrier and repositioned pilot subcarrier for 64 QAM modulation

In Fig. 6 a, b, c, d blue curves describe the variation of the bit error rate according to SNR that occurs when we add two pilot subcarriers. After compiling all plots, we can confirm that the green curves describing the rearranged pilot subcarrier impact yield better results than the standardized positions. We can also conclude that the new proposed approach offers the best values over all cases. By comparing the Fig 6.a and Fig 6.b with Fig 6.c and Fig 6.d, we can also conclude that the more we rise the data rate, the better results we obtain. Despite this enhancement, adding pilot subcarriers will affect the Spectral Efficiency SEF which is defined as the ratio between pilot subcarrier and the total number of used subcarriers. Indeed it increases from 0.076 to 0.115.

6. CONCLUSION:

802.11p is the IEEE standard that offers basic specification to the V2X communication in mobile Environments. Our research topic involves around studying the WAVE performance and its enhancement. We have focused our work on channel estimation algorithms and on performing pilot symbol estimation techniques. The main topic of this study is to combine the two approaches MMSE / LS estimation and pilot based estimation to enhance as better as possible the V2X communication performances. In a first step, we underlined the fact that MMSE estimator yields much better results than LS estimators. We have also proved that rearranging pilot subcarriers can offer better results than the standardized position. We have underlined in a second step that adding two new pilot subcarriers can enhance the performance of the V2X communication. We have also shown that appending more pilot subcarriers can affect the spectral efficiency: as the SEF grows from 0.076 to 0.115. We have come to reach the conclusion that although the fact that the proposed approach can be an enhancement solution, there remains a spectral inconvenience to be eventually overcome.